\documentstyle[aps,multicol,epsf]{revtex}
\begin{document}
\preprint{}
\draft
\title{Scaling near Quantum Chaos Border in Interacting Fermi Systems}
\author{Pil Hun Song\cite{adr}}
\address{Center for Theoretical Physics, Seoul National University, 
Seoul 151--742, Korea}
\date{\today}
\maketitle
\begin{abstract}
The emergence of quantum chaos for interacting Fermi systems is
investigated by numerical calculation of the level spacing distribution 
$P(s)$ as a function of interaction strength $U$ and the excitation energy 
$\epsilon$ above the Fermi level.  As $U$ increases, $P(s)$ undergoes 
a transition from Poissonian (nonchaotic) to Wigner-Dyson (chaotic) 
statistics and the transition is described by a single scaling parameter 
given by $Z = (U \epsilon^{\alpha}-u_0) \epsilon^{1/2\nu}$, where $u_0$
is a constant.  While the exponent $\alpha$, which determines the global 
change of the chaos border, is indecisive within 
a broad range of $0.9 \sim 2.0$, the finiteness of $\nu$, which comes from 
the increase of the Fock space size with $\epsilon$, suggests that the 
transition becomes sharp as $\epsilon$ increases.
\end{abstract}


\begin{multicols}{2}
\narrowtext

Recently, the emergence of quantum chaos due to electron-electron 
interaction has attracted much attention.  While the subject
has a longer history in nuclear physics, with the progress
of modern nanofabrication techniques, it has entered condensed 
matter physics since a quantum dot system, for example, could be 
considered as an artificial atom with some physical parameters 
under control.  Besides being of interest as its own right, the 
importance of the subject stems from the fact that it is related to 
a failure of the perturbative approach in interacting many-particle systems.  
Traditionally, the perturbative method has been one of the standard 
tools in theoretical many-particle physics.  The emergence of quantum chaos 
means a strong mixing of the unperturbed levels, thereby inducing 
breakdown of the perturbation series.  

A recent theoretical work by Altshuler {\it et al.}\cite{alt} for 
the quasiparticle decay in a quantum dot has especially stimulated 
many theoretical investigations. In their paper\cite{alt}, the 
quasiparticle decay process was mapped to a single particle diffusion 
on the Bethe lattice, which is a nonperturbative treatment of
the problem.  They concluded that there is a transition to quantum 
chaos at a critical excitation energy $\epsilon_c \sim \sqrt{g}\Delta$, 
where $g$ is the dimensionless conductance and $\Delta$ is the mean 
level spacing between the single-particle levels.  However, many 
authors afterwards have pointed out that the
mapping to a Bethe lattice in ref.~\cite{alt} is too naive for a proper 
description and they obtained different results using other methods.  
The ongoing controversy could be summarised through two main
questions.  i) what is the relation between $\epsilon$ and 
$g$ at the quantum chaos border, and ii) whether the 
transition is sharp or not.  Regarding the first question, there exist
further different predictions such as $\epsilon_c 
\sim g^{2/3}\Delta$\cite{jac,mir} and $\epsilon_c \sim g\Delta/\ln{g}$
\cite{sil}.  As for the second question, Jacquod and 
Shepelyansky\cite{jac} argued that the transition is smooth since 
the coupling between the Fock states is 
of nonlocal nature.  The authors of ref.~\cite{mej} concluded that
the transition is smooth based on their numerical result for
the local density of states and the participation number.  In ref.~\cite{ley},
the same conclusion has been drawn by calculating the inverse participation 
ratio (IPR) for higher values of $\epsilon$ by use of the, so called, layer
model.  On the other hand, Berkovits and Avishai\cite{ber} suggested a finite 
size scaling hypothesis, according to which the transition becomes sharp as
$\epsilon$ increases, which was based on their exact numerical results for 
small system size.  Silvestrov\cite{sil} also proposed that the transition 
is sharp when the effective high-order interaction is taken 
into account.  In addition, the authors of ref.\cite{ley2} performed
an analysis of their numerical result for the IPR and found that their
results show features consistent with the prediction of ref.~\cite{sil}. 
However, their conclusion is not decisive enough concerning the sharpness 
of the transition.  A reliable numerical test is, therefore, urgently 
needed to settle the issue.

The main difficulty of numerical test is due to the fact that one should 
consider 
the regime of $g \gg 1$, which corresponds to $\epsilon \gg \Delta$.  Since 
the size of the matrix to be diagonalized rapidly increases with $\epsilon$, 
one needs an alternative to the brute force method.  The layer model, 
introduced by Georgeot and Shepelyansky\cite{geo}, allows one to handle much 
larger system size (higher $\epsilon$) at a given computational cost 
by truncating out the Slater determinants which contribute little to 
a given eigenstate.

In this paper, we calculate the level spacing distribution $P(s)$ for
interacting fermionic systems up to $\epsilon/\Delta = 27$ by use of
the layer model.  The change of $P(s)$ from the Poissonian to Wigner-Dyson 
statistics
represents the transition from integrability to chaos.  While our result
does not allow us to resolve the controversy over the parametric relation
for the quantum chaos border, i.e. question (i), it gives
strong evidence that the transition becomes sharper as $\epsilon$
increases.  The finite size scaling (FSS) behaviour can be understood 
through a comparison with an infinite dimensional Anderson model.  

Let us begin with $n_f$ spinless fermions on $m$ single-particle levels for
which the Hamiltonian ${\cal H}$ is given by ${\cal H}_0 + {\cal H}_1$ with 
\begin{equation}
{\cal H}_0 = \sum_i \epsilon_i c^\dagger_i c_i \ \ \mbox{and} \ \ 
{\cal H}_1 = \sum_{i<j,k<l} U_{ij,kl} c^\dagger_l c^\dagger_k c_i c_j.
\end{equation}
${\cal H}_0$ represents the noninteracting Hamiltonian.  A single-particle 
level energy $\epsilon_i$ with $i = 1,2,\cdot\cdot\cdot, m$ is randomly 
chosen from the interval $[(i-1/2)\Delta,(i+1/2)\Delta]$ with uniform 
probability distribution.  $U_{ij,kl}$ denotes the random two-body 
interaction matrix element and also has a box distribution $[-U/2,U/2]$.  
When ${\cal H}$ is interpreted as the Hamiltonian for a quantum dot
system, $U$ is related to $g$ as $U \sim \Delta/g$\cite{alt,bla}.
${\cal H}$ is widely known as the Two-Body Random Interaction
Model (TBRIM) and its full matrix size is given by $N_t = n_f!/[m!(m-n_f)!]$.  
When ${\cal H}$ is written in matrix form on the basis of the 
eigenstates of ${\cal H}_0$, a diagonal element is the sum of the
energies of occupied single-particle levels, while an off-diagonal element is 
nonzero only for two states which are different upto two particle-hole 
pairs.  

Since $N_t$ increases very rapidly with $n_f$ and $m$, direct
numerical diagonalisation of ${\cal H}$ is limited to rather small
values of $n_f$ and $m$.  Due to this difficulty, the layer model 
has been introduced by Georgeot and Shepelyansky\cite{geo} and the 
details are as follows.
For each eigenstate of ${\cal H}_0$, we define a sequence $\{f_i\}$ 
such that $f_i$ = 0 when the level $i$ is empty and $f_i$ = 1 when 
it is filled with a fermion.
Then ${\cal E}/\Delta = \sum^{m}_{i=1} i f_i$ ranges from $n_f (n_f+1)/2$ 
to $n_f (m+1) - n_f (n_f+1)/2$, which is approximately equal to the
ground state energy and the highest excitation energy, respectively, in 
units of $\Delta$.  If we rewrite ${\cal H}$ in 
ascending order of ${\cal E}$ and set $U_{ij,kl} = 0$ for $i+j \neq
k+l$, the whole matrix reduces to a block diagonal form where
${\cal E}$ is constant within each block.  We call the submatrix 
with ${\cal E}/\Delta-n_f (n_f+1)/2 = j$ the $j$-th layer model.  
In general, the matrix size of the $j$-th layer model, $N(j)$, 
varies depending on $n_f$ and $m$.  However, for $1 \ll n_f \ll m$, 
$N(j)$ for $j\leq n_f$ is determined solely by $j$ and behaves as\cite{mot} 
\begin{equation}
N(j) \sim \exp(\pi \sqrt{2j/3})/j.
\end{equation}
The layer model hereafter is understood in such a sense.  An eigenstate 
of ${\cal H}$ with energy $E$ can be written as a superposition of the 
eigenstates of ${\cal H}_0$, the energies of which lie within the width 
$\Gamma$ around $E$.  $\Gamma$ is much less than $\Delta$, when $U \ll 
\Delta$, i.e. $g \gg 1$.  This defines the valid regime of the layer model.

One of the well established criteria for transition from 
integrability to chaos is the change of the level spacing distribution 
$P(s)$ from the Poissonian $P_p(s) = e^{-s}$ to the Wigner surmise 
$P_w(s) = (\pi s/2) e^{-\pi s^2/4}$.  To
quantify the proximity of $P(s)$ to either of the two, it is useful 
to define $\eta$ in terms of the variance of $s$ as follows;
\begin{equation}
\eta =\frac{\int_0^{\infty} s^2(P(s)-P_{w}(s)) ds}{\int_0^{\infty} 
s^2(P_{p}(s)-P_{w}(s)) ds}.
\end{equation}  
In this way, $\eta$ is 1 for $P_p(s)$, 0 for $P_w(s)$ and inbetween 
for an intermediate distribution.  

In our calculation, the Hamiltonian for the layer model for $19 \leq j 
\leq 27$ has been constructed and numerical diagonalisation has been 
performed over 200-1000 disorder configurations for each parameter set
$(j,g)$.  Corresponding matrix size ranges from 490 ($j=19$) to 3010
($j$=27).  To exclude the contribution of the tail states near the
edge, 50 $\%$ of eigenvalues around the band center have been used to 
construct $P(s)$.  If we choose a smaller part of the eigenvalue set, 
e.g. 25 $\%$, the result does not change significantly.

The $\eta$'s for $19 \leq j \leq 27$ and for $1/200 \leq U \leq 
1/10$\cite{com1} are shown in Fig.~1.  The gradual change from $\eta 
\simeq 1$ (Poissonian, integrable) to $\eta \simeq 0$ (Wigner-Dyson, 
chaotic) with increase 
of $U$ is found with all $j$.  As $j$ increases, the transition occurs 
at a smaller value of $U$, indicating that the interaction becomes 
more efficient for mixing the levels as the energy increases.  
The global dependence of the chaos border on the excitation energy may 
be found by plotting the data of Fig.~1 on a rescaled $x$-axis with 
$x = U j^{\alpha}$\cite{com2}.  If there exists $\alpha = \alpha_0$ such that
all the data points lie in the same curve, the transition is described
by the relation $\epsilon = U^{-1/\alpha_0}$ and there is a smooth crossover 
instead of a sharp phase transition when $\epsilon \rightarrow \infty$.
However there does not exist such a value $\alpha_0$ for
our data.  Instead, as shown in Fig.~2(a-c), there is a single crossing 
point where the curves meet one another and the slope at the crossing 
point becomes larger as $j$ increases.  This
suggests that at sufficiently high energy the transition to chaos is a
sharp phase transition.  In fact such a FSS feature 
with $j$ is found for a broad range of $0.9 \leq \alpha \leq 2$.  
When $\alpha$ is tuned from 0.9 to 2, $\eta_c$ ($\eta$ at the crossing 
point) decreases monotonically from 0.8 to 0.2.  For $\alpha < 0.9$
or $\alpha > 2$, the crossing point cannot be clearly identified.
These results imply that while we cannot determine the exponent
$\alpha$ which governs the quantum chaos border\cite{com3}, the 
transition shows a FSS property regardless of the 
choice of $\alpha$ as long as $0.2 < \eta_c < 0.8$.

For a further analysis of this FSS behaviour, we assume that 
$\eta$ is given by a function $f$ of a single scaling variable 
$Z$ such that
\begin{equation}
\eta(\epsilon,U) = f(Z) = f[(U \epsilon^{\alpha}-u_0) \epsilon^{1/2\nu}],
\end{equation}
where $u_0$ is a constant of the order of unity which is independent 
of $U$ and $\epsilon$.  To understand the introduction of this
scaling parameter $Z$, it would be helpful to compare the layer model 
with an Anderson Hamiltonian.  One can think of the standard Anderson 
Hamiltonian 
defined on a graph, which has the same structure as the $j$-th layer
model; there are $N(j)$ vertices and a bond between two vertices exists 
if two corresponding Slater determinants of the $j$-th layer 
model have nonzero transition amplitude by the two-body interaction.
The size of the graph can then be defined by the average minimum number 
of bonds between two vertices.  It is given by $\sim \sqrt{j}$, which 
is the same as the typical number of electron-hole pairs for the
$j$-th layer model\cite{mir}.  The spatial dimension 
of the graph in the limit of $j \rightarrow \infty$ is given by 
\begin{equation}
d=\lim_{j \rightarrow \infty} \frac{\ln N(j)}{\ln (\sqrt{j})}
\rightarrow \infty.
\end{equation}
We expect that there exists a transition to chaos as we decrease the
disorder strength of the on-site energy for a fixed value of hopping
energy between nearest neighbors. The transition will show a FSS behaviour 
with the system size $L \sim j^{1/2}$ and with $\nu =1/2$, the correlation 
length exponent of infinite dimensional Anderson model\cite{efe}.  
Returning back to the layer model, we expect that a similar FSS property 
exists since there also is the competition between the system size and 
a correlation length governing the chaotic property of the system.  
A difference between the layer models and the thus defined 
Anderson models is that the matrix elements are strongly correlated 
in the former, while they are completely random in the latter\cite{com4}.  
Therefore, $\nu$ of Eq.~(4) is not necessarily equal to that of the 
infinite Anderson model, 1/2.  In passing, we note that Berkovits and 
Avishai\cite{ber} introduced a scaling hypothesis very similar to Eq.~(4) 
based on their numerical results for lower energy.

According to Eq.~(4), the slope of the curves $\eta(U)$ for a given
$j$ behaves as $\sim j^{\alpha + 1/(2\nu)}$ at the crossing point and $\nu$ 
is obtained by fitting this variation of the slopes for a given value
of $\alpha$.  Fig.~3 shows the scaling plot obtained in this way with 
$\alpha = 1.5$ ($\eta_c = 0.56$) and in this case $\nu$ is given by 
0.59.  Such a procedure can
be performed in the same way for $0.2 \leq \eta_c \leq 0.8$ ($0.9 \leq \alpha 
\leq 2$) and the result for $1/(2\nu)$ is shown in Fig.~4.  Taking 
account of uncertainties of all data, $1/(2\nu)$ lies between 0.5 and 1.5 
($0.3 < \nu < 1$) over the whole range where the crossing point is 
identified.  Therefore, though $\nu$ also varies in a rather broad 
interval, our data evidently excludes the possibility of $\nu$ being
infinity, i.e. the possibility of a smooth crossover.  One should note 
that this large variation in $\nu$ over a factor of 3 is the result 
of the fact that $\alpha$ is indecisive.  If one can pin down the range 
of $\alpha$ in some other way than our method, or assume one of the 
previously proposed values (e.g. 1.5\cite{jac,mir} or 2\cite{alt}) to 
be valid, the uncertainty in $\nu$ can be determined to a higher 
accuracy from Fig.~4.

Finally, we discuss the sharpness of the transition when $\epsilon$ is
varied for a fixed value of $U$, since in real measurements with
a quantum dot the conductance of the sample $g \sim \Delta/U$ is kept constant
and the bias voltage, which corresponds to $\epsilon$, is varied.  We
define the transition as sharp when the ratio of the transition
interval $\delta \epsilon$ to the transition energy $\epsilon_c$ goes
to zero as $\epsilon_c$ increases.  From Eq.~(4), $\epsilon_c$ is
given by $(u_0/U)^{1/\alpha}$ and $\delta \epsilon$ is estimated
using the slope of the curve $\eta(\epsilon,U)$ as 
\begin{equation}
\delta \epsilon \sim \left| \frac{\partial\eta}{\partial\epsilon} 
\right|^{-1}_{\epsilon = \epsilon_c} \sim \ \ 
\epsilon_c^{1-\frac{1}{2\nu}}. 
\end{equation}
Therefore, $(\delta \epsilon/\epsilon)|_{\epsilon = \epsilon_c} 
\sim \epsilon_c^{-1/(2\nu)}$
and we find that the transition becomes sharp as long as $\nu$ is
not infinity.  Therefore, the result of Fig.~4 shows that the transition
to chaos takes place sharply if one uses a sufficiently clean sample,
i.e. large $g$, for a quantum dot.

In summary, we have numerically investigated the emergence of quantum
chaos in the interacting many-fermion systems using the layer model.
While our result does not allow us to find the global dependence of
the chaos border as a function of the interaction strength, we find that
the transition becomes sharp at sufficiently high energy.  This is
also true when one observes the transition to chaos in a quantum dot
as a function of bias voltage.

I am thankful to Jae Dong Noh, Gun Sang Jeon and especially to Dima 
L.~Shepelyansky for helpful discussions.

\pagebreak
\noindent
\begin{figure}
\centerline{\epsfxsize=8cm \epsfbox{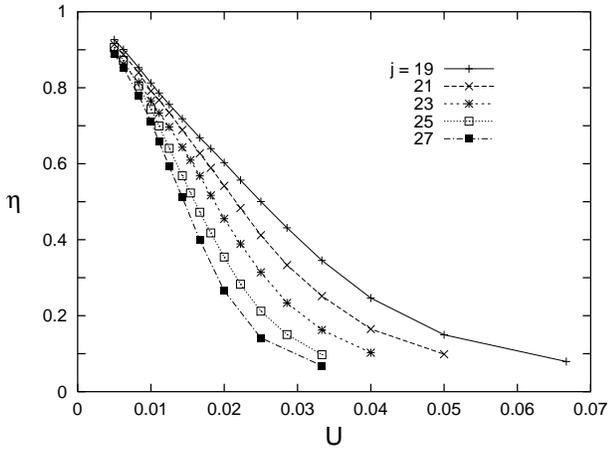}}
\vspace{4mm}
\caption{$\eta$ as a function of the interaction strength $U$ for
the layer model with various values of $j$.  
}
\end{figure}
\noindent
\begin{figure}
\centerline{\epsfxsize=6cm \epsfbox{fig2.epsi}}
\vspace{4mm}
\caption{The same data as Fig.~1 with rescaled $x$-axis of $x=U j^{\alpha}$ 
with $\alpha = 1.2$, 1.5 and 1.8.  A partial set of data from Fig.~1 is 
shown for clarity.
}
\end{figure}
\noindent
\begin{figure}
\centerline{\epsfxsize=8cm \epsfbox{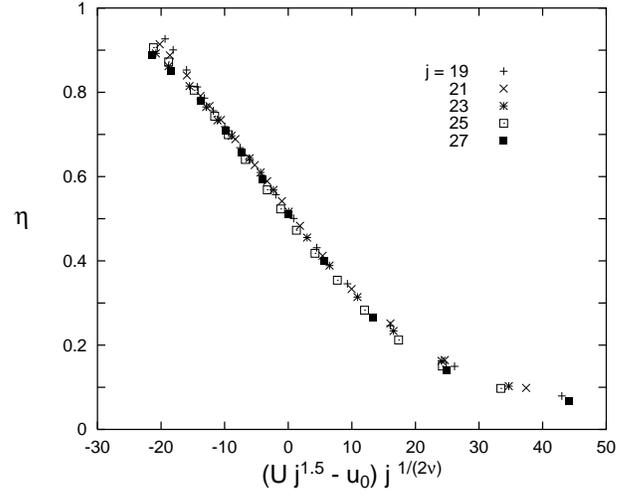}}
\vspace{4mm}
\caption{
Scaling plot of the data of Fig.~1 for $\alpha = 1.5$.  In
this case $1/(2\nu)$ is given by 0.85, i.e. $\nu = 0.59$.
}
\end{figure}
\noindent
\begin{figure}
\centerline{\epsfxsize=8cm \epsfbox{fig4.epsi}}
\vspace{4mm}
\caption{
$1/(2\nu)$ when $\eta_c$ varies from 0.2 to 0.8.
}
\end{figure}
\end{multicols}

\begin{thebibliography}{99}
\bibitem{com1}
We may set $\Delta=1$ without loss of generality.  Under this 
condition, the relations between parameters become simpler to yield 
$g \sim 1/U$ and $\epsilon=j$, for example.
\bibitem{com2}
We note that within the finite range of $1/200 \leq U \leq 1/10$ ($10 \leq
g \leq 200$), the functional form of $g/\log{g}$, which was suggested 
in ref.~\cite{sil}, does not differ seriously from a power form of 
$g^{1/\alpha}$.
\bibitem{com3}
We may try a more refined analysis through higher cumulants of $P(s)$
assuming that just at the transition point, $P(s)$ will be given by a
critical form $P_c(s)$ which is insensitive to $U$ and $j$.
Unfortunately, such an approach is not successful within our data
accuracy.
\bibitem{com4}
For example, the correlation of the diagonal energies for the
layer model decays very slowly with the distance as
$\langle {\cal H}_{ii} {\cal H}_{jj} \rangle \ \sim \ C-d(i,j)$,
where $C$ is a constant of order of $n_f$ and $d(i,j)$ is the distance
between two Slater determinants or two vertices characterized by $i$
and $j$ respectively.  In contrast, it is zero unless $i=j$ for the 
Anderson models.
\end{thebibliography}
\end{document}